# Tailoring Exchange Couplings in Magnetic Topological Insulator/Antiferromagnet Heterostructures

Qing Lin He[1]†, Xufeng Kou[1]†, Alexander J. Grutter[2], Lei Pan[1], Xiaoyu Che[1], Yuxiang Liu[1], Tianxiao Nie[1], Steven M. Disseler[2], Brian J. Kirby[2], William Ratcliff II[2], Qiming Shao[1], Koichi Murata[1], Yabin Fan[1], Mohammad Montazeri[1], Julie A. Borchers[2], and Kang L. Wang[1]*

[1]Department of Electrical Engineering, University of California, Los Angeles, California 90095, USA.

[2]NIST Center for Neutron Research, National Institute of Standards and Technology, Gaithersburg, MD 20899-6102, USA.

*Correspondence to: wang@ee.ucla.edu.

†These authors contributed to this work equally.

**Magnetic topological insulators such as Cr-doped $(Bi,Sb)_2Te_3$ provide a platform for the realization of versatile time-reversal symmetry-breaking physics. By constructing heterostructures with Néel order in an antiferromagnetic CrSb and magnetic topological order in Cr-doped $(Bi,Sb)_2Te_3$, we realize emergent interfacial magnetic phenomena which can be tailored through artificial structural engineering. Through deliberate geometrical design of heterostructures and superlattices, we demonstrate the use of antiferromagnetic exchange coupling in manipulating the magnetic properties of the topological surface massive Dirac fermions. This work provides a new framework on integrating topological insulators with antiferromagnetic materials and unveils new avenues towards dissipationless topological antiferromagnetic spintronics.**

In magnetic topological insulators (MTIs), time-reversal symmetry is broken and a gap is opened in the topological surface states by the ferromagnetic order, driving the original metallic surface Dirac fermions into a massive state[1-3] and imbuing the MTI with exotic time-reversal symmetry breaking physics including the quantum anomalous Hall effect[3], axion electrodynamics[4], and spin–orbit torque[5]. Incorporating magnetic elements into TIs has proven an effective approach to introduce ferromagnetism, as manifested by recent realizations[6-8] of the quantum anomalous Hall effect in Cr/V-doped $(Bi,Sb)_2Te_3$. The applicability of time-reversal-symmetry-breaking physics can be greatly expanded by utilizing magnetic interactions at high temperatures to increase the Curie Temperatures ($T_C$). For instance, it is proposed that when a TI is integrated with a high-$T_C$ ferromagnet, proximity effects at the interface involving the surface Dirac fermions will align spin moments of the TI band/itinerant carriers and give rise ferromagnetism[2]. Although many early attempts have been carried out[9-12], the induced magnetic spin polarization could not be extended through the entire structure, likely due to the short-range exchange coupling at the interface. Alternatively, by interfacing antiferromagnets (AFMs) with MTIs, it may be possible to realize more stable longer-range magnetic interactions. As an antiparallel magnetic structure with a low magnetic susceptibility, AFMs have negligible net magnetization[13,14]. Consequently, in a AFM/MTI heterostructure, the full spectrum of topological surface states may be easily modified through the strong exchange field of the AFM without invoking significant spin-dependent scattering on magnetic ions[15,16]. Furthermore, as AFM moments are relatively insensitive to external magnetic fields, the magnetism in such a heterostructure is resistant to external perturbations. Additionally, the interfacial coupling to pinned AFM spins gives rise to exchange bias, through which the MTI magnetization may be manipulated.

In this work, we incorporate AFMs into MTI heterostructures and utilize unique properties of AFMs to tailor the MTI magnetic ground state through inter- and intra-layer exchange couplings. Interactions between AFM spins and massive Dirac fermions within the MTI lead to a giant enhancement in the magnetic ordering of the MTI. The AFM CrSb (lattice constants a=4.122 Å)[17], is among the few antiferromagnetic materials that is lattice-matched with Cr-doped $(Bi,Sb)_2Te_3$ MTI (a=4.262-4.383 Å), making it an ideal candidate for the growth of epitaxial AFM/MTI heterostructures and superlattices (SLs) by molecular beam epitaxy (Fig. 1a). The magnetic moments of the MTI exhibit strong perpendicular anisotropy regardless of layer thickness[7,8] while in bulk AFM the Cr spins lie along the *c*-axis, exhibiting A-type AFM order with spins aligned ferromagnetically within the basal plane and antiferromagnetically between adjacent planes (Néel temperature around 700 K)[18]. To achieve highly insulating bulk character and enable surface Dirac fermion-mediated ferromagnetism in the MTI layer, the Bi: Sb ratio is optimized to be 0.26: 0.62 with various Cr doping concentrations, positioning the Fermi level within the surface gap[8].

We discovered novel interfacial magnetic interactions between the Dirac fermions and the AFM spins accompanied by a giant enhancement of magnetic ordering temperature (a factor of three increase of $T_C$) in the SL compared with a single MTI layer. The exchange coupling in the heterostructures is shown to originate from the Dirac fermions of MTIs and is responsible for a modified spin texture within the AFM relative to bulk CrSb. The important role of Dirac fermions is explored through the fabrications of different heterostructures, the control of Dirac fermion mass by using different Cr doping concentrations, and characterization by a range of experimental techniques including polarized neutron reflectivity.

The perpendicular anisotropy of the MTI layer (28 nm, Cr-doping concentration x=0.16), which is confirmed by the nearly square *M-H* loop obtained under an applied perpendicular magnetic field (Fig. 1b), is essential for opening a surface gap and achieving the quantum anomalous Hall effect. In contrast a negligible net magnetization is observed in the nominally AFM single layer (24 nm) of CrSb (Fig. 1b). In the bilayer AFM (24 nm)/MTI (28 nm) (Fig. 1c), we observed an increase of the coercive field ($H_C$)[19] to around 670 Oe for the bilayer (as compared with around 470 Oe for a MTI single layer) at 5 K. Such an increase is likely related to the exchange coupling between the Dirac fermions and the AFM spins at the interface, which is further evidenced by the interface-sensitivity: on one hand, by adding a second AFM layer (24 nm) at the bottom (*i.e.* forming an AFM/MTI/AFM trilayer, Fig. 1d), $H_C$ can be further enhanced to around 900 Oe given the fact that the number of the coupled interfaces is now doubled; on the other hand, the enhancement of $H_C$ is found to be gradually reduced by increasing the thickness of the MTI layer (Supplementary Information), indicating surface-dominated ferromagnetism. Equally important, the exchange bias of the bilayer resulting from couplings between the AFM spins with surface Dirac fermions give rise to a lateral shift of coercive fields under the perpendicular field-cooling condition, as observed in Fig. 1c and schematically illustrated by the red and blue curves in Fig. 1e, respectively[19]. The exchange bias persists up to around 35 K (Supplementary Information), slightly lower than $T_C$ (around 38 K) of a single MTI layer, indicating that the biased magnetization originates within the MTI. In addition to the lateral exchange-biased shift, we also observed a characteristic vertical shift in magnetization, which may originate from uncompensated AFM spins at the interface[20]. On the other hand, exchange bias is not observed in the symmetric AFM/MTI/AFM trilayer sample (Fig. 1d). All of these observations are consistent with the picture of exchange bias originating from a unidirectional perpendicular anisotropy established at the AFM/MTI interface[19], which implies the presence of strong interfacial exchange coupling established in the AFM/MTI bilayer via interactions between polarized Dirac fermions and the AFM spins.

Strikingly, if the AFM is instead sandwiched between two MTI layers, (*i.e.* MTI(28 nm)/AFM(24 nm)/MTI(28 nm) trilayer in Fig. 1g), a unique *M-H* loop and distinct magnetic interactions develop. In sharp contrast to the bilayer, such a trilayer shows a negligible exchange bias as the corresponding *M-H* loops after field-cooled or zero field-cooled treatments are essentially identical. Instead, a symmetrical *M-H* loop with a double switching feature arises potentially as a result of the interlayer exchange coupling, *i.e.,* coupling between two neighboring MTI layers. Interestingly, this interlayer exchange coupling is antiparallel and the spontaneous (near zero field) magnetizations of the two MTI layers are antiparallel. (Fig. 1f, shows a quantitative calculation of the resultant magnetic configurations arising from competitions between the Zeeman interaction, interfacial and internal exchange interaction, dipolar interaction and magnetic anisotropy, as discussed in Supplementary Information). Equally important, we note that exchange coupling between Dirac fermions and the AFM spins (dash circles in Fig. 1f) is crucial to stabilizing such a magnetic configuration, as supported by the single square-shape *M-H* loop with parallel MTI magnetization alignment in a control trilayer MTI/TI/MTI, in which a nonmagnetic TI layer $(Bi,Sb)_2Te_3$ layer replaces the AFM layer with the same thickness (Fig. 1h).

To quantitatively investigate the interplay between the exchange coupling of Dirac fermions and AFM spin configuration at the AFM/MTI interface, we further expand our structural engineering by using building blocks of AFM(3nm)/MTI(7nm) to construct *n*-period SLs over a range of *n*=2-15. The antiparallel interlayer exchange coupling, denoted by the symmetrical double switching behaviors in the corresponding *M-H* loops, is even more pronounced in all these SLs (Fig. 2a). More importantly, such coupling persists to a temperature that is much higher than $T_C$ of a single MTI layer. For example, $H_C$ of the *n*=4 SL (Fig. 2b) gradually decreases with increasing the temperature from 5 K to 70 K when the double switching becomes weakened; at temperatures above 80 K, the double switching disappears and is replaced by a paramagnetic response. Similar temperature dependences are also observed in SLs with *n*=2, 6, 8, 10, confirming the giant $T_C$ enhancement. Consistent with the MTI/AFM/MTI case discussed above, we also note that no signatures of exchange bias are observed in these SLs, again implying that the inter- or intra-layer exchange coupling exchange differs from that observed in the AFM/MTI bilayer. We thus explore the possibility that when sandwiched by two MTI layers, the spin texture of the AFM is modified from the bulk form, hindering the development of an unidirectional anisotropy in the MTI layer at the interfacial region.

To address the origin of such exotic interlayer magnetic interactions we utilized polarized neutron reflectometry (PNR) to extract and compare the detailed temperature-dependent magnetization profiles of an (AFM/MTI)$_{n=10}$ SL and the AFM/MTI/AFM trilayer. PNR was performed under an in-plane applied field of 700 mT to rotate the magnetization into the plane of the film. The fitted reflectivity of both the SL and trilayer, along with corresponding magnetic/structural depth profiles, are shown in Fig. 3. In the SL, we find a net magnetization of 47 and 62 emu/cm$^3$ on the MTI and AFM layers, respectively. Alternative models in which the net magnetizations are exclusively confined within either the MTI or AFM layers result in significantly worse fits to the data (Supplementary Information). It should be noted that in modeling the SL PNR spectra cannot well distinguish between a uniform and modulated magnetization profiles (*e.g.* Néel-type domain walls) within the AFM layers, and is consistent with a number of complex spin structures such as an oscillatory magnetization. On the other hand, PNR of the trilayer shows a magnetization that is mostly confined to the MTI layers. Models which assume a magnetization of 62 emu/cm$^3$ in the AFM layer fail to describe the data, while the best fit suggests a magnetization of 5 emu/cm$^3$ or less (Supplementary Information). At temperatures above $T_C$ of the MTI, no net magnetization on either layer is observed. Thus, PNR results demonstrate that the AFM develops a net magnetization only when sandwiched between two MTI layers, whereby the spin texture is altered from the bulk. This net magnetization is small, and, assuming full magnetic ordering of the AFM, represents a canting angle of approximately 16°.

Neutron diffraction measurements at zero applied field were carried out on a (AFM/MTI)$_{n=15}$ SL to provide additional insight into the AFM spin structure. The (0001) and (0002) hexagonal diffraction peaks are sensitive to the in-plane component of the antiferromagnetic and ferromagnetic structure respectively, and were probed between 5 K and 300 K. Given that the net magnetization represents a small fraction of the total Cr moments even in an applied field of 700 mT, we expect minimal magnetic intensity at the (0002) peak location. Indeed, diffraction measurements show exclusively structural scattering at this location (Supplementary Information). Furthermore, the structural (0001) peak is forbidden in CrSb, and any scattering observed in this location must be either be purely magnetic or originate in the MTI structure. However, the only peak observed near the CrSb (0001) is a temperature independent peak corresponding to the MTI (0006) reflection, and thus antiferromagnetically ordered spins may be oriented along *c*-axis in Fig. 1a. In such an orientation, the magnetization is parallel to $Q_Z$ and the neutrons are not sensitive to the magnetic moment. Alternatively, it is possible that the AFM order does not maintain phase coherence between AFM regions, resulting in magnetic peaks which are too broad to resolve. Combining the PNR and diffraction results, we conclude that the CrSb layers are magnetically ordered and that the spin structure is dramatically altered from the bulk due to the interlayer exchange coupling when sandwiched between two MTI layers (like the (AFM/MTI)$_n$ SLs and MTI/AFM/MTI cases). Consequently, exchange bias eliminates. In contrast, for an AFM layer in contact with a single MTI layer (like the AFM/MTI bilayer and the AFM/MTI/AFM trilayer cases), the AFM spin texture remains mostly intact, and exchange bias mediated by the AFM spins and Dirac fermions develops.

Based on the observations of a strong interfacial exchange coupling and antiparallel interlayer exchange coupling, we suggest a Néel-type domain wall extending into the interiors of the AFMs in the MTI/AFM/MTI trilayer and SLs in an applied field. Quantitative simulations and calculations, which include the competing effects of interfacial exchange and Zeeman energy (Supplementary Information), visualize possible spin texture through the heterostructure as shown in Fig. 1f. Accordingly, the observed strong interfacial exchange coupling between the Dirac fermions and the AFM spins demonstrate a high spin-transparency and magnetic coherence of the AFM/MTI interface (note that PNR reveals a weak ferromagnetic moment within AFM layers which could facilitate the interlayer exchange coupling), and therefore, such a Néel-type magnetic domain wall configuration is believed to result from the exchange coupling between the neighboring MTI layers mediated through the AFM layer. Consequently, long range effects on the magnetization in the SLs are established.

In the following, we show that such a long-range exchange coupling along the *c*-axis direction enables a modification of the magnetic ordering in the entire SLs and an enhancement of the ordering temperature of the MTI layers. The onsets of ferromagnetisms in both the MTI/AFM/MTI trilayers and AFM/MTI bilayers are found to appear at significantly higher temperatures than the single MTI layer with the same Cr-doping concentration (*e.g.,* x=0.16, the corresponding $T_C$s are 58, 54, and 39 K, respectively). Additionally, we observe 90% spontaneous magnetization in the (AFM/MTI)$_{n=4}$ SL (Supplementary Information), further demonstrating the nearly single domain nature[6] of the SL, which is in sharp contrast to the single MTI layer. These observations show that both the interfacial and interlayer exchange couplings contribute to the enhancement of magnetic ordering. Inspired by these results, we used minor loop measurements (Fig. 2c) to extract the bias field due to interlayer exchange coupling ($H_{EX}$), which serve as an approximation of the coupling strength[21-23]. Figures 4a and b show the temperature-dependent $H_{EX}$ of *n*-period SLs and the corresponding isotherms. $H_{EX}$ increases dramatically with SL period up to *n* = 4, reaching a plateau of around 45 mT when *n* > 4, implying an optimized interlayer coupling length scale ($L_{IEC}$) along the *c*-axis direction which characterizes the maximum coupling energy. The maximum $L_{IEC}$ for (AFM/MTI)$_n$ SLs is therefore the distance between the top and bottom MTI layers of a *n*=4 SL (around 23 nm). It should be highlighted that the coupling along the *c*-axis direction crosses many van der Waals gaps within MTI quintuple-layers, showing signatures of long-range Dirac-fermion-mediated ferromagnetism[24-26]. Accordingly, $T_C$ as a function of *n* (Fig. 4c) is extracted through Arrott plots of Hall resistances (Supplementary Information), and it can be clearly seen that the increase of $T_C$ saturates when *n* ≥ 4. In addition to $H_{EX}$ and $T_C$, the enhancement is also demonstrated by the increased $H_C$ (Fig. 4e). In comparison, a single MTI layer does not show a distinct thickness-dependent $H_C$, implying dominant carrier-independent Van Vleck magnetism[27]. In short, the consistency of $n$- and temperature-dependencies of the magnetic parameters ($H_{EX}$, $T_C$, and $H_C$) highlights roles of both interfacial and interlayer exchange coupling in magnetic ordering enhancement. To further confirm that the long range magnetic order may be attributed to Dirac fermions coupled to AFM spins, Cr doping concentration was varied in order to adjust the magnitude of the gap (and thus mass) of the surface states of the MTI. The increased mass of Dirac fermions with increased Cr-doping will decrease the spin-orbit coupling strength and eventually eliminate the nontrivial bulk band topology. This in turn will tune the ferromagnetic exchange (*i.e.*, with a higher Cr doping level, the surface gap is increased, and the Dirac fermion mediated ferromagnetism is weakened)[28]. Experimentally $H_{EX}$ in the SLs decreases with increasing $M_s$ (corresponding to increasing Cr concentrations x=0.05, 0.09, 0.13, 0.16, and 0.19) respectively (Figs. 4d), consistent with the Dirac-fermion-mediated ferromagnetism.

These results show that the long-range interlayer exchange coupling likely originates from exchange between the polarized Dirac fermions in different MTI layers, mediated and amplified by coupling to the AFM spins[29]. The interaction is expected to be maximized when coupling between MTI layers is established throughout the entire SL, leading to a saturation of the magnetic ordering enhancement. This behavior also suggests that AFMs may act as excellent exchange-interaction transmitters with Dirac fermions.

We have demonstrated intimate connections between heterostructure geometry and magnetic spin textures of the MTIs and AFMs. The AFM is shown to be an efficient interfacial- and interlayer-exchange coupling mediator with Dirac fermions of the MTI, which additionally allows a giant enhancement in magnetic ordering and a modification of the composite magnetic order.

Particularly we demonstrated several advantages of AFM-based proximity effect engineering relative to ferromagnet-based systems, including longer-range interactions. This work unveils enormous opportunities for integrating and modifying topological surface states through coupling to AFM order, opening new avenues towards structural engineering of MTIs and showcasing the benefits of integrating MTIs with AFMs.

## Methods:

(a) MBE growth of hybrid magnetic structures

The MBE growths of all the magnetic systems in this work were fabricated in an ultra-high vacuum Perkin-Elmer MBE system. The growth of Cr-doped $(Bi,Sb)_2Te_3$ thin films were reported previously by the same group while the CrSb thin films were fabricated in the same growth chamber. All the samples presented in this study were fabricated on epi-ready semi-insulating GaAs(111)B substrates. Prior to sample growth, Se-assisted oxide-desorption processes were carried out at 580 °C for 30 minutes to form thin GaSe buffer layers. High-purity Bi and Cr were evaporated from standard Knudsen cells while Sb, Se, and Te were evaporated by standard thermal cracker cells. The temperatures of the substrates were then decrease to around 200°C during the growths. A real-time reflection high energy electron diffraction (RHEED) was used to monitor all the growth cycles. During the growths of both Cr-doped $(Bi,Sb)_2Te_3$ and CrSb thin films, the RHEED patterns were optimized to very sharp, smooth, streaky patterns, while the intensity oscillations were used to calibrate the growth rate. We noticed that a direct epitaxial growth of CrSb on the GaAs substrate results in three dimensional growth mode with low crystalline quality. Thus, the growth of a CrSb single layer is carried out on an undoped 2nm-thick $(Bi,Sb)_2Te_3$ buffer layer on the GaAs substrate, through which the crystalline quality of CrSb becomes much more satisfactory. The growth rates of Cr-doped $(Bi,Sb)_2Te_3$ and CrSb thin films were determined to be around 0.08 and 0.06 Å/s respectively. After the films' growths, 2 nm of Al layers were evaporated to all the sample surfaces *in situ* at room temperature for protection against contamination and oxidation.

(b) Polarized Neutron Reflectometry

Polarized neutron reflectometry (PNR) measurements were carried out with the PBR beamline at the NIST Center for Neutron Research. Samples were field cooled in an in-plane applied field of 700 mT to a temperature of 20 K. Measurements were performed in the specular reflection geometry, with the direction of wave vector transfer perpendicular to the superlattice surface. The neutron propagation direction was perpendicular to both the sample surface and the applied field direction. Based on magnetometry measurements showing in-plane magnetization saturation below the applied field of 700 mT, spin-flip scattering is not expected. Therefore, we measured the spin-up and spin-down reflectivies using full polarization analysis to ensure that the incident and scattered beams retained identical neutron polarization directions. We refer, therefore, only to the spin-up and spin-down non-spin-flip reflectivities, which are a function of the nuclear and magnetic scattering length density profiles. The magnetic profile was deduced through modeling of the data with the NIST Refl1d software package.

(c) Neutron Diffraction

Neutron diffraction measurements were performed on the BT-4 triple axis spectrometer and polarized beam measurements were performed on the BT-7 triple axis spectrometer with $^3$He spin filters at the NIST Center for Neutron Research. Measurements were carried out in a temperature range of 5-300 K in a closed-cycle refrigerator. The incident and scattering neutron energy was 14.7 meV ($\lambda$ = 2.359 Å), selected by Pyrolitic graphite (PG) monochromator and analyzer crystals with multiple PG filters before and after the sample to eliminate higher order neutrons. The Soller collimator configuration in downstream order was open-monochromator-40'-sample-40'-analyzer-open-detector. For polarized beam measurements two different spin quantization axes were measured independently; the first with vertical spins out of the scattering plane and therefore perpendicular to the (0002), and the second with spins polarized horizontally along the momentum transfer *Q*.

(d) Magneto transport measurement

Hall bar devices with dimensions of 2 mm × 1 mm were fabricated for the transport measurements in a Quantum Design physical property measurement system. Systematically altering experimental variables such as temperature, magnetic field, working frequency, and rotation angle, in addition to multiple lock-in-amplifiers, and Keithley source meters enable comprehensive and high-sensitivity transport measurements in all the devices.

**Acknowledgments:**

This work was supported as part of the SHINES Center, an Energy Frontier Research Center funded by the U.S. Department of Energy, Office of Science, Basic Energy Sciences under Award # S000686. We are grateful to the support from the ARO program under contract 15-1-10561. We also acknowledge the support from the FAME Center, one of six centres of STARnet, a Semiconductor Research Corporation program sponsored by MARCO and DARPA. Certain commercial equipment, instruments, or materials are identified in this paper to foster understanding. Such identification does not imply recommendation or endorsement by the National Institute of Standards and Technology, nor does it imply that the materials or equipment identified are necessarily the best available for the purpose.

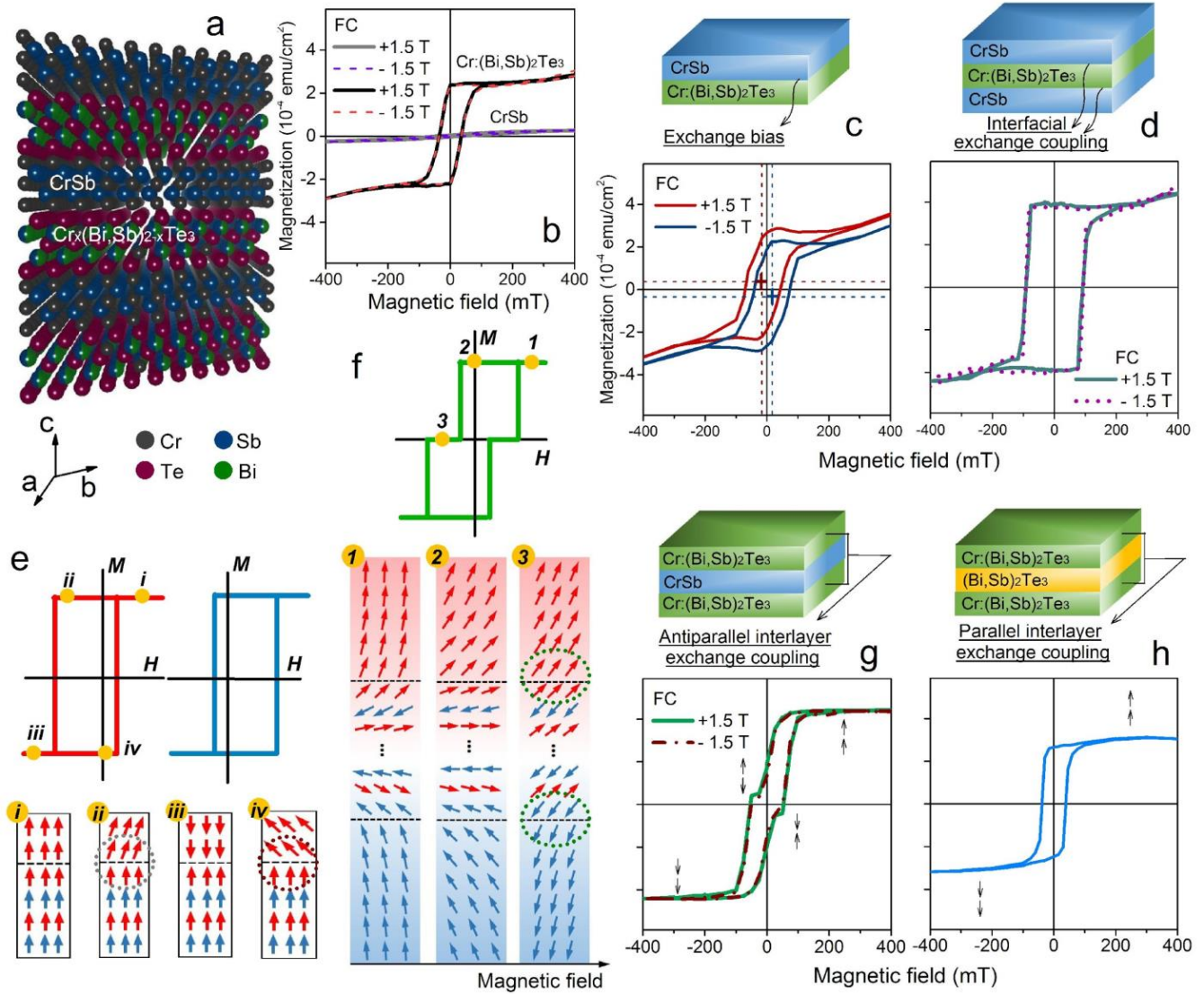

**Fig. 1. Exchange couplings of Dirac fermions and AFM for different heterostructures. a**, A schematic building block of an AFM (CrSb) / MTI [Cr-doped $(Bi,Sb)_2Te_3$] superlattice used in the work. **b**, The perpendicular anisotropy of a MTI single layer is demonstrated by the square-shaped *M-H* loops measured under a perpendicular magnetic field at 5 K, while the AFM order of the AFM single layer is evidenced by the negligible net magnetization. **c**, Exchange bias observed in an AFM/MTI bilayer is demonstrated by the lateral shift (marked by dash lines and stars) of the *M-H* loops (red and blue) under different perpendicular field-coolings. Cooling in a positive field

results in a negative shift (red loop, -108 Oe) and vice versa (blue loop, +110 Oe). **d**, In an AFM/MTI/AFM trilayer, the two coupled AFM/MTI interfaces dramatically increase the coercive field of the MTI but eliminate the exchange bias due to the symmetry of the structure, whose effects from two surfaces are canceled out. **e**, The schematic orientations of the atomic moments in the exchange-biased AFM/MTI bilayer, illustrating strong magnetic interactions between Dirac fermions and AFM spins. When the external field reverses the magnetization of the MTI, the AFM spins do not follow the field but try to keep the FM spins in their original orientations. Consequently the external field needed to reverse an exchange-biased MTI is larger than for a single MTI layer. **f**, in a MTI/AFM/MTI trilayer, the relative orientations of the atomic moments of the antiparallel interlayer exchange coupling, which is generated from a quantitative simulation and calculation based on the results of polarized neutron reflectometry described in Fig. 3. The interfacial exchange coupling between the Dirac fermions and the AFM spins orients the magnetization directions of the two MTI layers, forming a stable field-induced Néel-type domain wall within the AFM. **g**, A novel antiparallel interlayer exchange coupling is observed in a MTI/AFM/MTI trilayer, which supports that the idea that this texture is mediated by the interactions between Dirac fermions and AFM spins. **h**, A control trilayer that replaces the AFM layer in **g** by an undoped TI layer with the same thickness, which is nonmagnetic. In this system, the interlayer exchange coupling is eliminated; this result demonstrates the significant role of the interactions between Dirac fermions and AFM spins for stabilizing the magnetic configuration.

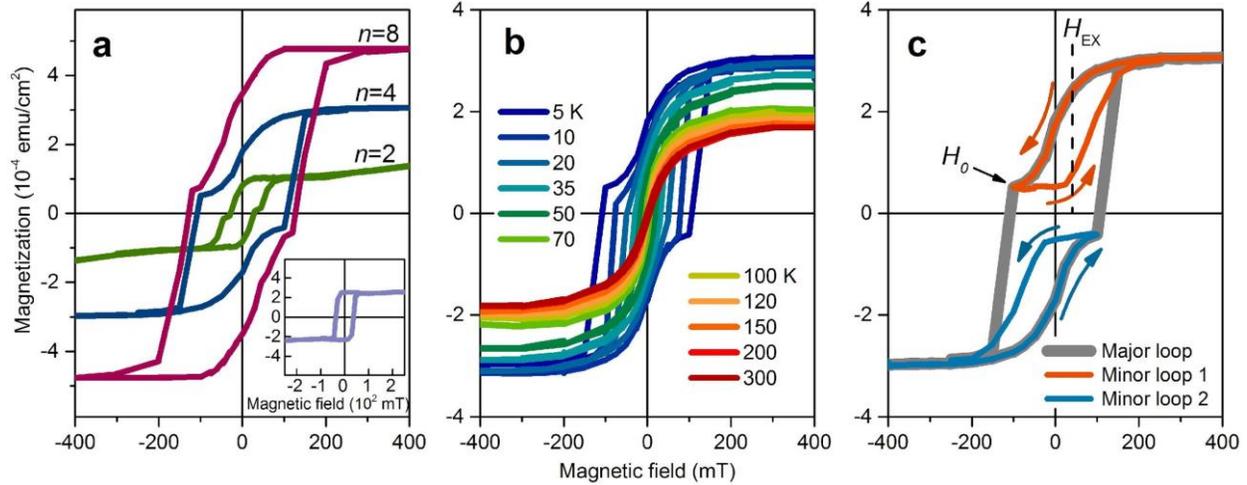

**Fig. 2. Novel magnetic interplays in MTI/AFM superlattices (SLs). a**, Interlayer exchange couplings in (AFM/MTI)$_n$ SLs with $n$=2, 4, and 8, as demonstrated by the double-step behaviors in the *M-H* loops measured at 5 K. Similar to the trilayer in Fig. 1**g**, this behavior is attributed to the antiparallel magnetizations between the neighboring MTI layers in the SLs, which originates from the interactions between Dirac fermions and AFM spins. The inset shows the magnetization of a control SL (TI/MTI)$_{n=4}$, in which the AFM layers are replaced by undoped TI layer with the same thickness. In contrast, the lack of exchange couplings between Dirac fermions and AFM spins only contributes a single step switching behavior. As shown in **b**, the double-step switching behavior persists up to ~ 80 K, after which it becomes paramagnetic. **c**, Exchange field $H_{EX}$ between the neighboring MTI layers in the SLs probed by minor loop measurements, as marked by the dash line. The grey loop indicates the major loop, while the red-orange and blue loops indicate the two minor loops. A typical minor loop 1(2) begins at +(−)1.5 T, decreases to −(+) a specific value ($H_0$), then sweeps back to +(−) 1.5 T; $H_0$ is a critical field value that is just above the onset of the second step switching.

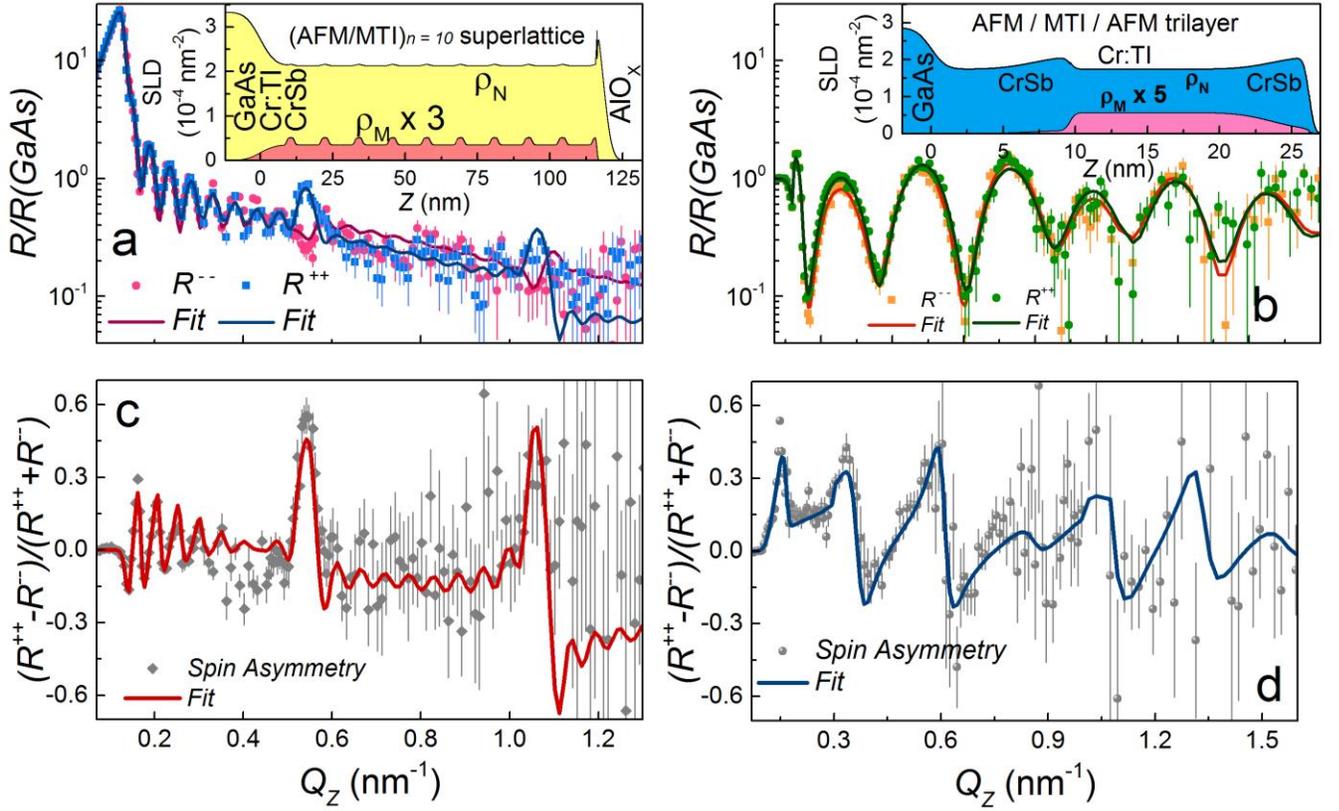

**Fig. 3. Capturing the spin textures in the SL and trilayer by neutron techniques.** Polarized neutron reflectivities (at 20 K with a 700 mT in-plane field) normalized to the GaAs substrates for the spin-polarized $R^{++}$ and $R^{--}$ channels of **a**, an (AFM/MTI)$_{n=10}$ SL and **b**, an AFM/MTI/AFM trilayer. The insets show the corresponding models with structural and magnetic scattering length densities (SLDs), $\rho_N$ and $\rho_M$, used to obtain the best fits, which contain magnetized TI and AFM layers for the SL, and magnetized TI and barely magnetized AFM layers for the trilayer. In the SL, the partial magnetizations within the AFM layers are likely due to the interlayer exchange coupling with Dirac fermions from the neighboring MTI layers, which demonstrates that the AFM is acting as a mediator to propagate the exchange interactions. In contrast, in the trilayer the interaction between Dirac fermions and the AFM spins at the interface dominates and the AFM texture remains almost intact. The detailed spin asymmetry $(R^{++}- R^{--})/(R^{++}+ R^{--})$ between the $R^{++}$ and $R^{--}$

channels for **c**, the SL and **d**, the trilayer. In **c**, only both the MTI and AFM layers are magnetized can well fit the spin splitting, while the magnetization is assumed to be confined exclusively to the MTI layers or the AFM layers, *i.e.* fail to describe the signs of spin-splitting. In **d**, the best fit suggests that the AFM layer is barely magnetized (5 emu/cc or less, intrinsic AFM nature) while models assuming magnetized AFM layers fail to describe the spin asymmetry. The error bars are +/- 1 standard deviation.

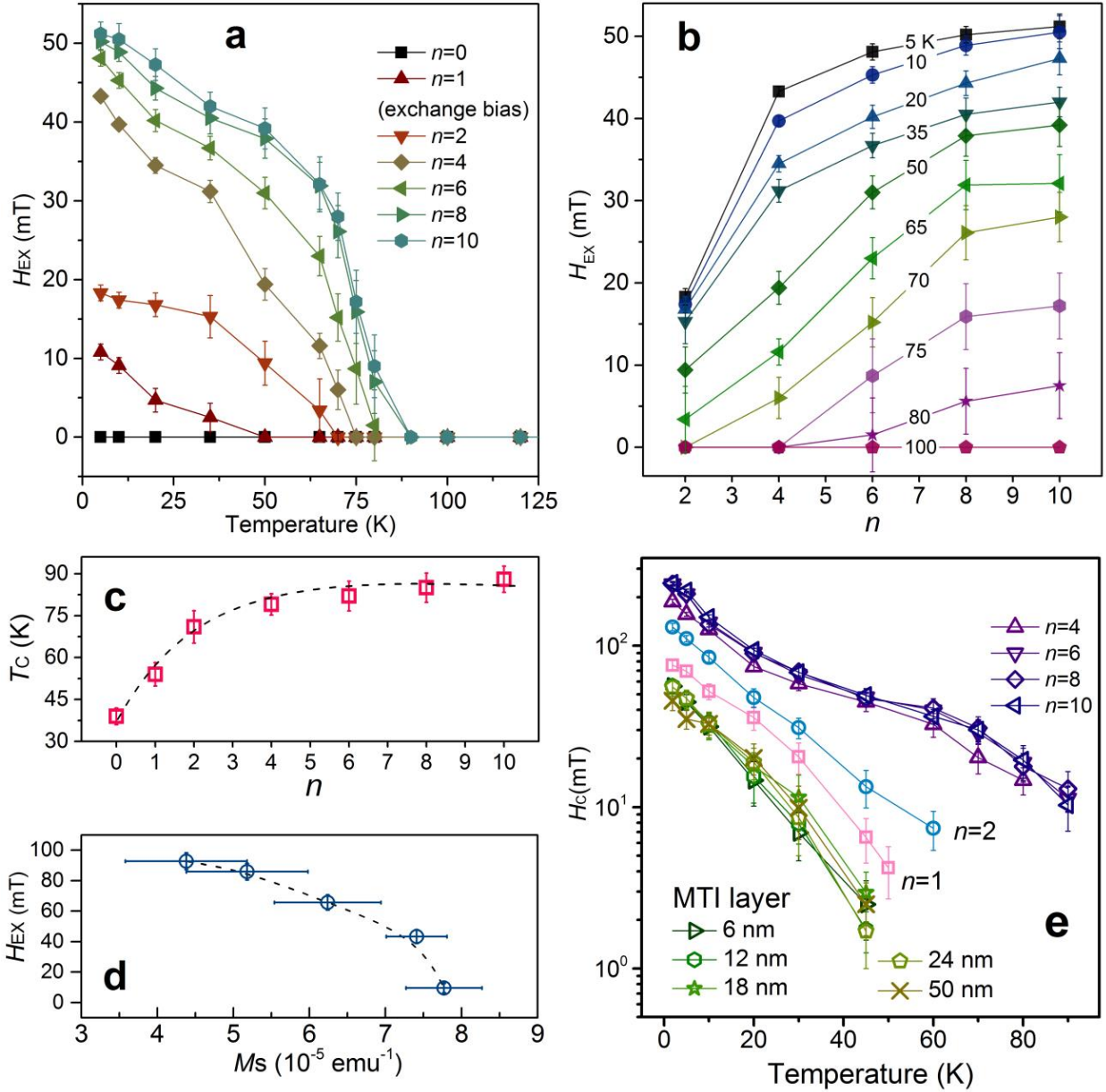

**Fig. 4. Observation of giant enhancements in exchange field ($H_{EX}$), Curie temperature ($T_C$), and coercive field ($H_C$) in the superlattices. a**, Temperature-dependent $H_{EX}$ obtained from (AFM/MTI)$_n$ SLs. It is noticed that $H_{EX}$ persists at temperatures much higher than $T_C$ of the MTI single layer, implying that the enhancement of magnetic ordering is strongly correlated with the interlayer exchange coupling. **b**, $H_{EX}$ isotherms as a function of $n$, which clearly shows the saturation of interlayer exchange coupling when $n \geq 4$. Such saturation behavior can also be observed in **c**, $T_C$ *vs* $n$. **d**, $H_{EX}$ as a function of $M_S$, which corresponds to different Cr-doping concentrations x=0.05, 0.09, 0.13, 0.16, and 0.19 in five (AFM/MTI)$_{n=4}$ SLs, indicating the diminishing exchange coupling between the AFMs and the Dirac fermions. The decreases of $H_{EX}$ along with the increasing $M_S$ demonstrates the weakening of interlayer exchange couplings between MTI layers, implying a heavy Cr doping concentration increases the acquisition of a mass term in the topological surface states. **e**, $H_C$ of (AFM/MTI)$_n$ SLs *vs T*, showing $H_C$ increases along with the increase of $n$, consistent with correlation between the interlayer exchange coupling and magnetic ordering enhancement.